\documentclass[12pt,nofootinbib,showpacs,aps]{revtex4}
\usepackage{graphics}
\begin{document}

\title{The eternal fractal in the universe}

\author{Serge Winitzki}
\affiliation{Department of Physics and Astronomy, Tufts University,
Medford, MA 02155, USA}

\begin{abstract}
Models of eternal inflation predict a stochastic self-similar geometry
of the universe at very large scales and allow existence of points
that never thermalize. I explore the fractal geometry of the resulting
spacetime, using coordinate-independent quantities. The formalism
of stochastic inflation can be used to obtain the fractal dimension
of the set of eternally inflating points (the ``eternal fractal'').
I also derive a nonlinear branching diffusion equation describing
global properties of the eternal set and the probability to realize
eternal inflation. I show gauge invariance of the condition for presence
of eternal inflation. Finally, I consider the question of whether
all thermalized regions merge into one connected domain. Fractal dimension
of the eternal set provides a (weak) sufficient condition for merging.
\end{abstract}

\maketitle 

\section{Introduction}

In models of cosmological inflation, amplification of quantum fluctuations
of the scalar field (the inflaton) may result in eternal self-reproduction
of inflating domains, or \textit{eternal inflation}
\cite{Vil83,EIall,EIambig,EImods}. In this case, stochastic evolution
produces a spacetime in which thermalization never happens globally. At
arbitrarily late times there exist vast regions of space where inflation
still continues; however, along any given comoving world-line,
thermalization is certain to be reached (with probability $1$) at a
sufficiently late proper time. This somewhat paradoxical situation is
possible when self-reproduction of inflating regions proceeds faster than
the drift of the inflaton field toward thermalization within these regions.

To study the spacetime resulting from eternal inflation, one may use
a stochastic formalism \cite{Vil83,Starob,FPEall,LLM94} that describes
the probability distribution of the inflaton field $\phi $, coarse-grained
in horizon-size volumes of space. In simplest one-field models of
inflation, the time-dependent distribution of physical volume $P_{V}\left(
\phi ,t\right) $ satisfies a Fokker-Planck (FP) equation
\begin{equation}
\label{eq:FP-phys}
\frac{\partial P_{V}}{\partial t}=\frac{\partial ^{2}}{\partial \phi
^{2}}\left( DP_{V}\right) -\frac{\partial }{\partial \phi }\left(
vP_{V}\right) +3HP_{V}.
\end{equation}
This equation%
\footnote{Here and below we use the Ito factor ordering of the diffusion
operator \cite{Vil99FO}.
} is supplemented by appropriate initial and boundary conditions; $D\left(
\phi \right) $ and $v\left( \phi \right) $ are diffusion and drift
coefficients and $H\left( \phi \right) $ is the effective expansion rate.
The kinetic coefficients are related to the inflaton potential $V\left(
\phi \right) $ by
\begin{equation}
\label{eq:HDv}
H\left( \phi \right) =\sqrt{\frac{8\pi V\left( \phi \right) }{3}},\:
D=\frac{H^{3}}{8\pi ^{2}},\: v=-\frac{1}{4\pi }\frac{dH}{d\phi }.
\end{equation}
At late times, solutions of Eq.~(\ref{eq:FP-phys}) are dominated
by the eigenfunction $\psi _{V}\left( \phi \right) $ of the diffusion
operator with the largest eigenvalue $\gamma _{V}$,
\begin{equation}
\label{eq:FPsol}
P_{V}\left( \phi ,t\right) \sim e^{\gamma _{V}t}\psi _{V}\left( \phi
\right) .
\end{equation}
If $\gamma _{V}>0$, the total volume of the inflating domain
$\int P_{V}\left( \phi ,t\right) d\phi $ grows exponentially
with time, which indicates presence of eternal inflation.

One problem with the present form of the stochastic formalism is its
dependence on the choice of equal-time surfaces. By construction,
the volume distribution at constant time $P_{V}\left( \phi ,t\right) $
is not a generally covariant quantity, and it is no surprise that
solutions and eigenvalues of the FP equation depend non-trivially
on the choice of the time variable $t$. The resulting spacetime
is highly inhomogeneous on large scales, and different choices of
time slicing introduce significant biases into the volume distribution.
For this reason, it has been difficult to draw unambiguous conclusions
about eternally inflating spacetimes using stochastic formalism
\cite{LLM94,EIambig,WV96}. One could even question the meaning of the
condition for presence of eternal inflation, $\gamma _{V}>0$, since the
eigenvalue $\gamma _{V}$ is itself gauge-dependent.

Eternal self-reproduction of inflating regions gives rise to fractal
structure of eternally inflating spacetime on very large scales.
Previously, the fractal structure of the inflating domain has been
investigated using distributions of $\phi $ on equal-time surfaces
\cite{AV87,Vil92,LLM94}. However, the fractal dimension $d_{fr}$ of the
inflating domain obtained in this way is determined by an eigenvalue of the
FP equation and depends, in general, on the choice of time parametrization
(except for the case when $H\left( \phi \right) \equiv \textrm{const}$).

Another open issue concerns the global topology of the eternally
self-reproducing universe: do thermalized domains remain forever separated
from each other by an inflating sea, or do all thermalized domains
eventually merge into one connected domain surrounding inflating islands?
If merging occurs, all other thermalized domains in the universe will
eventually enter our horizon; otherwise, we shall remain forever causally
separated from other thermalized domains by eternally inflating walls.
An early work by Guth and Weinberg \cite{GW83} showed that in the
old inflationary scenario there is no percolation of thermalized bubbles.
The same occurs in certain models of topological inflation \cite{TopInf}
with inflating domain walls. Numerical simulations of Linde
\textit{et~al.}~\cite{LLM94,EIglob} suggest that thermalized regions merge
in some chaotic inflationary scenarios. However, a conclusive general
description of the topology of the thermalized domain is lacking.

In the present work, I consider models of inflation that can be described
by Eq.~(\ref{eq:FP-phys}) or its generalizations for multiple scalar
fields. I demonstrate that certain global properties of eternally
inflating spacetimes can be determined in a coordinate-independent
way. The principal object under consideration is the ``eternal
fractal set'' $E$, the set of all comoving points that never
thermalize. The set $E$ is defined (Sec.~\ref{sec:ef}) independently
of the choice of coordinates. I show (Sec.~\ref{sec:fdFP}) that
its fractal dimension, $\textrm{dim}E$, is a gauge-invariant
quantity which coincides with the fractal dimension $d_{fr}$
of the inflating domain computed using the scale factor time variable.
In Sec.~\ref{sec:cond-ef} I study conditions for presence of eternal
inflation. First, I show that the basic criterion, $\gamma _{V}>0$,
is in fact gauge-invariant. Then I define the probability $X\left( \phi
\right) $ to have eternal points in a given comoving domain; this
probability plays the role of density of points of $E$. The quantity
$X\left( \phi \right) $ is gauge-independent and satisfies a nonlinear
branching diffusion equation derived in Sec.~\ref{sec:Xequ}. This new
equation is shown (Sec.~\ref{sec:cons}) to provide a (sufficient) condition
for presence of eternal inflation that is consistent with the first
criterion $\gamma _{V}>0$. Finally, in Sec.~\ref{sec:global} I consider the
question of whether all thermalized domains ultimately merge and become
connected to each other at sufficiently late times. I show that the
probability for merging is constrained by the nonlinear diffusion equation
of Sec.~\ref{sec:Xequ}. A (weak) sufficient condition for merging,
$\textrm{dim}E<2$, is provided by the fractal dimension of the set $E$. I
conjecture that the fractal dimension may provide a stronger criterion of
merging of thermalized domains. I also generalize the results to a suitable
class of global topological properties of the set $E$.

\section{The eternal fractal}

\label{sec:ef}
In the current picture of eternal inflation, certain regions of spacetime
may expand by an arbitrary factor and, in particular, arbitrarily small
sub-Planckian scales may become macroscopically large. There is evidence
that spacetime cannot be considered classical below the Planck scale (see
Ref.~\cite{Garay} for a review). Recently, it was shown that a
hypothetical modification of physics on sub-Planckian scales may change
the power spectrum of inflationary perturbations \cite{TP01} and create a
significant backreaction affecting the evolution \cite{TP02}. In that
case, the kinetic coefficients of Eq.~(\ref{eq:FP-phys}) would have to be
modified. However, the qualitative features of an eternally
self-reproducing spacetime would remain the same. In the present work we
shall assume that no new physics emerges from previously sub-Planckian
scales in the course of inflationary self-reproduction.

Let us consider the evolution of an arbitrary finite (comoving)
volume of space chosen at some initial time, in a cosmological model
with eternal inflation. Here and below we assume that the initial
comoving volume at $t=0$ had a certain value $\phi _{0}$
of the inflaton field and was physically of one horizon size $H^{-1}\left(
\phi _{0}\right) $.

There is a non-vanishing, if small, probability that the entire comoving
region will thermalize at a certain finite time, because fluctuations
of the scalar field $\phi $ could have accidentally cooperated
to drive $\phi $ toward the end point of inflation $\phi _{*}$
everywhere in the region. One could say in that case that eternal
inflation has not been realized in that region.%
\footnote{Here we do not consider the possibility of spontaneous return of
thermalized regions to inflation in ``recycling universe'' scenarios
\cite{GVrec}. } If, on the other hand, eternal inflation has been realized,
then for an arbitrarily late time $t$ there will be regions that are
still inflating at that time.

In that case, there must exist some comoving world-lines that never
enter a thermalized domain. To demonstrate this, one could choose
a monotonically increasing sequence of time instances, $t_{n}=nH^{-1}$,
$n=0,1,2,...$; for each $n$ there must exist a point $x_{n}$
surrounded by a horizon-sized domain that will be still inflating
at time $t_{n}$. (As $t_{n}$ grows, these domains become
progressively smaller in comoving coordinates, due to expansion of
space.) An infinite sequence of points $x_{n}$ on a finite (compact)
comoving volume of $3$-dimensional space must have accumulation
points, \textit{i.e.}~there must be at least one point $x_{*}$
such that any arbitrarily small (comoving) neighborhood of $x_{*}$
contains infinitely many points $x_{n}$ from the sequence. It
is clear that the comoving world-line at an accumulation point $x_{*}$
cannot reach thermalization at any finite time: if it did, there would
exist a comoving neighborhood around $x_{*}$ which thermalized
at that time, and this contradicts the construction of the point $x_{*}$.
We shall refer to points $x_{*}$ that never thermalize as ``eternal
points'' and define the set $E$ of all such points, as a set
drawn on the spatial section of the comoving volume at initial time.
(One could also imagine an infinitely dense grid of comoving world-lines
starting at the initial surface, with the set $E$ consisting
of all world-lines that never reach thermalization.) The set $E$
for a comoving region is not empty if eternal inflation has been realized
in that region.

Different histories and different choices of the initial comoving
volume will generate different sets $E$. Since $E$ is a
stochastically generated set, one may characterize it probabilistically,
by finding probabilities for the set $E$ to have certain properties.
The underlying stochastic process is the random walk of the inflaton
$\phi $ and, since it is a stationary stochastic process that
does not explicitly depend on time, probabilities of any properties
of the set $E$ depend only on $\phi _{0}$. For instance,
below we shall denote by $X\left( \phi _{0}\right) $ the probability
for the set $E$ to be non-empty if the initial comoving volume
has horizon size and has a given value $\phi =\phi _{0}$ of the
inflaton field.

During stochastic evolution, there will be (infinitely many) times
when the inflaton returns to the value $\phi _{0}$ in some horizon-sized
domain; each time the distribution of properties of the subset of
$E$ within those domains will be the same as that for the whole
set $E$. In this way, the set $E$ naturally acquires fractal
structure.

\subsection{The fractal dimension}

The comoving volume $P\left( \phi ,t\right) $ of inflating regions
with field value $\phi $ at time $t$ is described by the
FP equation
\begin{equation}
\label{eq:FP-c}
\frac{\partial P}{\partial t}=\frac{\partial ^{2}}{\partial \phi
^{2}}\left( D\left( \phi \right) P\right) -\frac{\partial }{\partial \phi
}\left( v\left( \phi \right) P\right)
\end{equation}
with the kinetic coefficients from Eq.~(\ref{eq:HDv}). Its late-time
asymptotic solution is
\begin{equation}
P\left( \phi ,t\right) \sim e^{-\gamma t}\, \psi \left( \phi \right) .
\end{equation}
The total comoving volume of inflating regions $\int P\left( \phi ,t\right)
d\phi $ decays exponentially rapidly at late times, and any pre-selected
comoving worldline eventually reaches thermalization with probability $1$.
In other words, the comoving volume occupied by eternal points vanishes
and the eternal set $E$ has measure zero, as a subset of 3-dimensional
(comoving) space. Its fractal dimension must therefore be less than
three.

The fractal set $E$ can be thought of as an infinite-time limit
of the (fractal) inflating domain considered in
Refs.~\cite{AV87,Vil92,LLM94}. Since the volume of $E$ vanishes, we need a
slightly different definition of fractal dimension than that of
Ref.~\cite{AV87}. The definition of fractal dimension we would like to use
is the so-called ``box fractal dimension'', which in our case is
equivalent to the Hausdorff-Besicovitch fractal dimension \cite{Feder}. The
``box fractal dimension'' is defined for a non-empty subset $S$ of
a finite region of a $d$-dimensional space. Rectangular coordinates
are chosen in the region to divide it into a regular lattice of
$d$-dimensional cubes with side length $l$. (The cubical shape of the
lattice is not essential to the definition of the ``box fractal
dimension''.) Then, the ``coarse-grained volume'' $V\left( l\right) $ of
the set $S$ is defined as the total volume of all cubes of the
lattice that contain at least one point from the set $S$. If
the set $S$ has fractal dimension less than $d$, one expects
$V\left( l\right) $ to decay at $l\rightarrow 0$ as some
power of $l$. The fractal dimension is defined by
\begin{equation}
\label{eq:fd-def}
\textrm{dim}S=d-\lim _{l\rightarrow 0}\frac{\ln V\left( l\right) }{\ln l}.
\end{equation}
For example, if the set $S$ is made of a finite number of line
segments, then the coarse-grained volume at scale $l$ will be
equal to the volume of narrow tubes of width $l$ surrounding
the line segments. Then $V\left( l\right) \sim l^{d-1}$ and the
fractal dimension defined by Eq.~(\ref{eq:fd-def}) would come out
to be $1$.

Below we shall see that if the eternal set $E$ is not empty,
it has a well-defined fractal dimension $\textrm{dim}E$ which
is independent of the choice of spacetime coordinates, and we shall
give a procedure to compute $\textrm{dim}E$ in a given inflationary
model.

\subsection{Topology of the eternal fractal}

Some simple topological properties of the eternal set $E$ follow
directly from its construction.

The set $E$ is topologically closed: if a point $x$ is not
an eternal point, $x\! \notin \! E$, then there exists a neighborhood
of $x$ which also does not belong to $E$. This is so because
the point $x$ must have thermalized at a finite time $t$
together with a horizon-size region around it.

The set $E$ does not contain isolated points: in any neighborhood
of a point $x\! \in \! E$ there exists, with probability $1$,
another point from $E$. To show this, suppose that a neighborhood
of a point $x\! \in \! E$ contains no other points of $E$.
Define $t_{0}$ to be the time when this neighborhood grows to
horizon (or larger) size and let $H_{0}$ be the Hubble constant
at point $x$ at that time. Consider an infinite sequence of horizon-sized
inflating regions $R_{n}$ surrounding the point $x$ at times
$t_{n}=t_{0}+nH_{0}^{-1}$, $n=0,1,2,...$; by assumption,
the evolution of the region $R_{n}$ at each $e$-folding
is such that $R_{n}$ expands into $e^{3}$ inflating subdomains,
of which one is to become $R_{n+1}$ and all others contain no
eternal points. There is a certain probability $p<1$ for a horizon-sized
region $R_{n}$ containing one eternal point to have such an evolution
during one $e$-folding; but the probability for \textit{all} regions
$R_{n}$, $n=0,1,2...$ to have the same evolution is the
product of infinitely many factors $p$, which is zero.

By extension, any neighborhood of $x\! \in \! E$ contains, with
probability $1$, infinitely many other points from $E$.

Since $E$ has measure zero, any given comoving point has \textit{a
priori} zero probability to belong to $E$. Therefore, any fixed
line segment or surface has zero probability to consist entirely of
points from $E$.

\subsection{Gauge invariance of fractal dimension}

Previous calculations \cite{AV87,Vil92,LLM94} concerned the fractal
dimension of the inflating domain at a finite time and assumed a certain
time slicing (gauge). The results depended on the choice of the time
coordinate (except for the case of Ref.~\cite{Vil92} where expansion
of inflating domains was homogeneous and all choices of time coordinate
are equivalent). Here, instead of the set of points that are inflating
at a given time, we consider the set $E$ of eternal points as
a subset of comoving spatial section at initial time. Now we shall
show that the fractal dimension of the set $E$ is invariant under
smooth changes of spacetime coordinates.

The set $E$ can be thought of as a locus of eternal points drawn
in comoving spatial coordinates on the initial space-like slice (at
$t=0$). The construction of the set $E$ requires to know
whether a given point has thermalized at \textit{any} finite time $t>0$.
Therefore the eternal set does not depend at all on the choice of
time slicing of the spacetime to the future of the original slice.
Its properties may, at most, be functions of the initial value of
$\phi $ at $t=0$.

Further, one can show that the fractal dimension of a set drawn in
a finite region of space remains unchanged under any smooth change
of coordinates in that region. First, the fractal dimension of a set
is unchanged under a linear transformation of coordinates with a
non-degenerate Jacobian $J$, because the coarse-grained volume $V\left(
l\right) $ will simply gain a factor $J$ and this will not change the
fractal dimension defined by Eq.~(\ref{eq:fd-def}). Given an arbitrary
smooth coordinate transformation, we can choose a sufficiently small scale
$l_{0}$ at which the coordinates are locally changed by a constant
linear transformation. We can divide the initial region into subdomains
of size $l_{1}<l_{0}$ and find $V\left( l\right) $ for $l<l_{1}$
by adding the volumes $V_{i}\left( l\right) $ calculated over
these subdomains. Under the coordinate change, each volume $V_{i}\left(
l\right) $ will be multiplied by a corresponding Jacobian $J_{i}$, but the
Jacobian in a given subdomain does not depend on $l$ for sufficiently
small $l_{1}$. Therefore, in the limit $l\rightarrow 0$
the dominant dependence of $V\left( l\right) $ on $l$ of
the form $V\left( l\right) \sim l^{\gamma }$ will be unchanged
under coordinate transformations.

Finally, the fractal dimension of the eternal set $E$ is the
same when computed in an arbitrarily small neighborhood of any eternal
point. This is because the set $E$ has no isolated points and
any point $x\! \in \! E$ is surrounded by infinitely many other
eternal points lying arbitrarily close to $x$. The statement
can be justified more formally as follows. In the next subsection
we show (without using the present statement) that the fractal dimension
of $E$ is, in fact, independent of the initial value of $\phi $
in the initial horizon-sized region. If we choose a small comoving
neighborhood $U\left( x\right) $ of an eternal point $x$,
there will be a time $t_{1}$ when $U\left( x\right) $ grows
to horizon size. Since $U\left( x\right) $ contains an eternal
point $x$, it could not have thermalized by the time $t_{1}$.
Therefore at time $t_{1}$ the neighborhood $U\left( x\right) $
is undergoing inflation and it can be taken as the initial comoving
region with a certain value of $\phi $. The future evolution
of any inflating horizon-sized region is statistically the same and
therefore the fractal dimension of the subset of $E$ computed
in the neighborhood $U\left( x\right) $ starting at $t=t_{1}$
is the same as that of the full set $E$.

Unlike other fractals occurring in nature, the fractal set $E$
has no short-distance cutoff: it retains its fractal structure at
all scales (in comoving space).

\subsection{Fractal dimension from Fokker-Planck equation}

\label{sec:fdFP}In this section we show that the fractal dimension
of the eternal set $E$ can be obtained from the comoving-volume
probability distribution of the field $\phi $ in a special gauge.
This probability distribution is a solution of the comoving-volume
FP equation in that gauge. The dominant eigenvalue of this FP equation
determines the fractal dimension of the set $E$ and coincides
with the dominant eigenvalue of the FP equation in the scale factor
time variable.

We introduce a special time variable $\theta $ related to the
scale factor $a$ by
\begin{equation}
\theta \left( t,{\mathbf x}\right) =\ln a\left( t,{\mathbf x}\right) +\ln
\frac{H\left( t,{\mathbf x}\right) }{H_{0}}.
\end{equation}
The meaning of this variable is the physical expansion scale $a\left(
t,{\mathbf x}\right) $ relative to the local horizon scale $H^{-1}$; a
surface of equal $\theta $ consists of spacetime points $\left( t,{\mathbf
x}\right) $ where the initial horizon size $H_{0}^{-1}$ had expanded to
$e^{\theta }$ times the current horizon size at the same point. One may
call $\theta $ a ``horizon scale factor'' time variable.

The time variable $\theta $ is not strictly monotonic: $\dot{\theta
}=\ddot{a}/\dot{a}$ (dots denote proper time derivatives) and if $H$ is
decreased rapidly enough so that the accelerated expansion locally stops,
the ``time'' $\theta $ will decrease despite growth of the scale
factor $a$. However, this is expected to happen only near the
end of inflation, whereas in other regimes such large fluctuations
of $H$ are extremely rare and expansion is almost always accelerated.
We shall assume that the time variable $\theta $ is well-behaved
throughout the allowed interval of $\phi $ and truncate that
interval near the end of inflation if necessary.

Below we shall see that in the gauge $\theta $, the comoving
volume probability distribution $P^{(\theta )}\left( \theta ,\phi \right) $
at late times $\theta $ has the form
\begin{equation}
\label{eq:ptheta-asymp}
P^{(\theta )}\left( \theta ,\phi \right) \approx \psi ^{(\theta )}\left(
\phi \right) e^{-\gamma \theta },
\end{equation}
where $(-\gamma )$ is the largest exponent dominating the solution
at late times and $\psi ^{(\theta )}$ is an appropriate eigenfunction.
Assuming that Eq.~(\ref{eq:ptheta-asymp}) gives the asymptotic solution
at late times, one finds the fractal dimension of the set $E$
to be
\begin{equation}
\label{eq:fd-ei}
\textrm{dim}E=3-\gamma .
\end{equation}

This result can be derived from Eq.~(\ref{eq:ptheta-asymp}) as follows.
Let $d=3$ be the dimension of space. An initial comoving region
of horizon size with a constant field value $\phi _{0}$ has linear
extent $H_{0}^{-1}$, where $H_{0}=H\left( \phi _{0}\right) $,
and we can choose comoving coordinates in the region to divide it
into $l^{-d}$ identical cubical subdomains of equal initial linear
size $lH_{0}^{-1}$, with $l\ll 1$. At a sufficiently late
time $\theta _{1}\left( l\right) $ when these comoving subdomains
have grown to horizon size, there will be on average $l^{-d}P^{(\theta
)}\left( \theta _{1},\phi \right) d\phi $ subdomains still inflating with
field value within the interval $[\phi ,\phi +d\phi ]$. (We assume that
eternal inflation has been realized in the initial region, otherwise the
number of inflating subdomains would be zero at sufficiently late times.)
The probability for a horizon-size region with field value $\phi $ to
contain at least one eternal point is a function of $\phi $ which we denote
$X\left( \phi \right) $. It follows that at time $\theta _{1}$ there will
be, on average, $l^{-d}\int P^{(\theta )}\left( \theta _{1},\phi \right)
X\left( \phi \right) d\phi $ subdomains that contain an eternal point.
Since $e^{\theta _{1}}=l^{-1}$ and $\theta _{1}$ is large for small $l$, we
can use Eq.~(\ref{eq:ptheta-asymp}) to obtain the total (comoving) volume
of those subdomains,
\begin{equation}
\label{eq:v-equ}
V\left( l\right) =e^{-\gamma \theta _{1}}\int \psi ^{(\theta )}\left( \phi
\right) X\left( \phi \right) d\phi =\textrm{const}\cdot l^{\gamma }.
\end{equation}
This is the coarse-grained volume of the eternal set at scale $l$.
By definition of fractal dimension, $\textrm{dim}E=d-\gamma $.

The constant in Eq.~(\ref{eq:v-equ}) is nonzero if $X\left( \phi \right)
\not \equiv 0$ and if the set $E$ is not empty; otherwise we cannot assume
that there are many inflating subdomains and estimate $V\left( l\right) $
by its average value, as we have done in Eq.~(\ref{eq:v-equ}). This
constant also absorbs all dependence on the initial value of $\phi $
in the initial comoving region. It follows that the fractal dimension
is independent of the initial value of $\phi $ and is well-defined
as long as eternal inflation is realized in the domain of consideration.

Now we need to determine $\gamma $. Instead of using a Fokker-Planck
equation in the gauge $\theta $, we will show that $\gamma $
is equal to the dominant eigenvalue of the Fokker-Planck equation
in the scale factor time variable. This will give a computational
prescription to obtain the fractal dimension of the set $E$ in
a given model of inflation.

The distribution of volume $P^{(\theta )}\left( \theta ,\phi \right) $
in the time gauge $\theta $ is related to the distribution $P\left(
t^{(0)},\phi \right) $ in the scale factor time gauge $t^{(0)}\equiv \ln a$
by
\begin{equation}
\label{eq:ptrel}
P\left( t^{(0)},\phi \right) d\phi =P^{(\theta )}\left( t^{(0)}+\ln
\frac{H\left( \phi \right) }{H_{0}},\phi \right) d\phi .
\end{equation}
To justify this, consider the combined comoving volume of all domains
in which the value of the field $\phi $ is between $\phi _{1}$
and $\phi _{1}+d\phi $ and scale factor is $\exp [t^{(0)}]$.
This volume is $P(t^{(0)},\phi _{1})d\phi $. {[}The distribution
$P(t^{(0)},\phi )$ as obtained from the FP equation is not normalized
since $\int P(t^{(0)},\phi )d\phi <1$ is the total comoving volume
of inflating regions in units of the initial comoving volume.{]} Throughout
all volume characterized by $\phi _{1}$ and $t^{(0)}$, the
``horizon scale factor'' $\theta $ has the value $\theta (t^{(0)},\phi
_{1})=t^{(0)}+\ln [H(\phi _{1})/H_{0}]+O\left( d\phi \right) $. Also, any
volume in which $\theta =\theta (t^{(0)},\phi _{1})$ and $\phi \in \left[
\phi _{1},\phi _{1}+d\phi \right] $ must have scale factor $\exp
[t^{(0)}]+O\left( d\phi \right) $. Therefore, $P(t^{(0)},\phi _{1})d\phi $
is, up to $O\left( d\phi ^{2}\right) $, equal to the total volume of all
domains in which $\theta =\theta (t^{(0)},\phi _{1})$ and $\phi \in \left[
\phi _{1},\phi _{1}+d\phi \right] $. This justifies Eq.~(\ref{eq:ptrel}).

The same justification can be given for an analogous relation between
physical-volume distributions in gauges $\theta $ and $t^{(0)}$,
since the argument does not rely on the comoving nature of the volume
to which the distributions relate.

The FP equation for the comoving volume distribution in the scale
factor time gauge $t^{(0)}$ is
\begin{equation}
\label{eq:FPt}
\frac{\partial P}{\partial t^{(0)}}=\frac{\partial ^{2}}{\partial \phi
^{2}}\left( D^{(0)}P\right) -\frac{\partial }{\partial \phi }\left(
v^{(0)}P\right) .
\end{equation}
Here
\begin{equation}
\label{eq:Dv-t0}
D^{(0)}=\frac{H^{2}}{8\pi ^{2}},\quad v^{(0)}=-\frac{1}{4\pi
H}\frac{dH}{d\phi }
\end{equation}
are the kinetic coefficients in this gauge. At late times $t^{(0)}$,
the solution of Eq.~(\ref{eq:FPt}) is
\begin{equation}
P\left( t^{(0)},\phi \right) \approx e^{-\gamma t^{(0)}}\psi \left( \phi
\right) ,\quad t^{(0)}\gg 1,
\end{equation}
where $\gamma $ is the dominant eigenvalue of Eq.~(\ref{eq:FPt}).
Late times $t^{(0)}$ correspond also to large values of $\theta $
since $H/H_{0}$ is bounded. It follows from Eq.~(\ref{eq:ptrel})
that the distribution $P^{(\theta )}$ has the asymptotic form
\begin{equation}
P^{(\theta )}\left( \theta ,\phi \right) \approx e^{-\gamma \theta }\left(
\frac{H\left( \phi \right) }{H_{0}}\right) ^{-\gamma }\psi \left( \phi
\right) ,\quad \theta \gg 1.
\end{equation}

Therefore, the dominant eigenvalue $\gamma $ of the FP equation
in the scale factor time gauge is the relevant quantity for the fractal
dimension in Eq.~(\ref{eq:fd-ei}).

This result shows that the fractal dimension of the set $E$ given
by Eq.~(\ref{eq:fd-ei}) is the same as the fractal dimension $d_{fr}$
of inflating domain \cite{AV87,Vil92,LLM94} if the latter is computed
on surfaces of equal scale factor. Although $\gamma $ is an eigenvalue
of the FP equation in a particular gauge, the quantity $d-\gamma $
has a physical interpretation as the fractal dimension of a set defined
independently of spacetime coordinates; eigenvalues of the FP equation
in other gauges do not seem to have a gauge-invariant interpretation.

\section{Conditions for eternal inflation}

\label{sec:cond-ef}We have distinguished two issues regarding the
possibility of eternal inflation: first, eternal inflation may be
entirely disallowed in certain models; and second, in models where
eternal inflation is possible, it may accidentally not have realized
in certain domains. In this section, we first examine conditions on
an inflationary model which make eternal inflation possible, and then
we shall obtain the probability for eternal inflation to be realized
in a given comoving region. The condition that the latter probability
does not vanish is another criterion for presence of eternal inflation.
Both criteria are shown to be independent of the choice of spacetime
coordinates. Finally, we demonstrate that the two criteria for presence
of eternal inflation are consistent with each other.

\subsection{Presence of eternal inflation}

The basic condition for possibility of eternal inflation in an inflationary
model is $\gamma _{V}>0$, where $\gamma _{V}$ is the dominant
eigenvalue of Eq.~(\ref{eq:FP-phys}). However, this equation itself
and the value of $\gamma _{V}$ are gauge-dependent. We shall
now show that $\gamma _{V}>0$ is in fact a gauge-invariant condition.

The key idea of the proof is that if $\gamma _{V}=0$ in some
gauge it must also be equal to zero in all other gauges, and therefore
$\gamma _{V}$ cannot be positive in one gauge and negative in
another.

Suppose that there exist two time variables, $t_{0}$ and $t_{1}$
such that the dominant eigenvalue $\gamma _{V}$ has opposite
sign in the gauges defined by $t_{0}$ and $t_{1}$. Introduce
a one-parametric family of time variables $t_{\alpha }$, $0\leq \alpha \leq
1$ to interpolate between these time variables,
\begin{equation}
\label{eq:timea}
\frac{dt_{\alpha }}{dt_{0}}\equiv \left( \frac{dt_{1}}{dt_{0}}\right)
^{\alpha }.
\end{equation}
The new time variables $t_{\alpha }$ are defined by integrating
Eq.~(\ref{eq:timea}) along comoving world-lines. The dominant eigenvalue
$\gamma _{V}$ of the FP equation in gauge $t_{\alpha }$
is a function of $\alpha $. This function $\gamma _{V}\left( \alpha \right)
$ is continuous, because the differential operator in the FP equation
is self-adjoint, its dominant eigenvalue $\gamma _{V}$ is non-degenerate
\cite{WV96} and under a small change of $\alpha $ this eigenvalue
must change by a small amount, found from standard perturbation theory.
By continuity it follows that there exists a value $\alpha _{0}$
such that $\gamma _{V}\left( \alpha _{0}\right) =0$. The FP equation
in this time parametrization has a stationary solution $P_{V}^{(0)}$,
\begin{equation}
\label{eq:FPsta}
\frac{\partial ^{2}}{\partial \phi ^{2}}\left( D_{\alpha
}P_{V}^{(0)}\right) -\frac{\partial }{\partial \phi }\left( v_{\alpha
}P_{V}^{(0)}\right) +3H_{\alpha }P^{(0)}_{V}=0.
\end{equation}
Here $D_{\alpha }$, $v_{\alpha }$ and $H_{\alpha }$
are kinetic coefficients in the gauge $t_{\alpha }$. They differ
from $D_{0}\left( \phi \right) $, $v_{0}\left( \phi \right) $
and $H_{0}\left( \phi \right) $ in the gauge $t_{0}$ by
the factor $\left( dt_{\alpha }/dt_{0}\right) ^{-1}$. The factor
$dt_{\alpha }/dt_{0}$ can be absorbed into $P_{V}^{(0)}$,
and we obtain a stationary solution of the FP equation in all other
gauges $t_{\alpha }$. Existence of this solution with eigenvalue
$0$ in all gauges $t_{\alpha }$ contradicts our assumption
that the largest eigenvalue of the FP equation is negative in one
of the gauges $t_{0}$ or $t_{1}$.

We have found that the dominant eigenvalue $\gamma _{V}$ of
Eq.~(\ref{eq:FP-phys}) must have the same sign in all gauges and,
therefore, presence or absence of eternal inflation can be judged by the
sign of $\gamma _{V}$ in any gauge.

\subsection{A diffusion equation for $ X\left( \phi \right) $ }

\label{sec:Xequ}Assuming now that $\gamma _{V}>0$, we consider
the probability for eternal inflation to be realized in a given region.
In Sec.~\ref{sec:fdFP}, we have denoted by $X\left( \phi \right) $
the probability for an initial inflating region of one horizon size,
in which the scalar field has value $\phi $, to contain at least
one eternal point. It will be convenient to work with the complementary
probability $\bar{X}\left( \phi \right) =1-X\left( \phi \right) $.
In this section we find that $\bar{X}\left( \phi \right) $ satisfies
a nonlinear branching diffusion equation {[}Eq.~(\ref{eq:Xequ}){]}
that resembles the backward FP equation for the physical volume probability
distribution. We derive this equation and show its independence of
time parametrization.

Here we shall use proper time $t$ as the time variable. The random
walk of the inflaton field $\phi $ is described by a Langevin
equation which can be written as a difference equation \cite{Starob},
\begin{equation}
\label{eq:lang}
\phi \left( t+\Delta t\right) =\phi \left( t\right) +v\left( \phi \right)
\Delta t+\xi \sqrt{2D\left( \phi \right) \Delta t},
\end{equation}
where $\xi =\xi \left( t,{\mathbf x}\right) $ is a normalized
random ``noise'' representing fluctuations, and $v\left( \phi \right) $,
$D\left( \phi \right) $ are the kinetic coefficients given by
Eq.~(\ref{eq:HDv}). The evolution of $\phi $ in a horizon-size
domain with field value $\phi =\phi _{0}$ at time $t=0$
gives rise to independent random walks of $\phi $ in each of
the ``daughter'' horizon-size subdomains that were formed out
of the original domain. After one $e$-folding, \textit{i.e.}~after
time $\Delta t=H^{-1}$, there are $N\equiv \exp \left( 3H\Delta t\right)
\approx 20$ daughter subdomains of approximately horizon size. The
stochastic process corresponding to $\xi $ assigns a probability density
$P\left( \xi _{1},...,\xi _{N}\right) d\xi _{1}...d\xi _{N}$
for various sets of values of $\xi $ in the $N$ daughter
subdomains.

The probability $\bar{X}\left( \phi _{0}\right) $ to have no
eternal points in the region is equal to the probability to have no
eternal points in any of its $N$ daughter subdomains. The evolution
in inflating daughter subdomains proceeds independently, so the probability
to have no eternal points in each of them is described by the same
function $\bar{X}\left( \phi \right) $ evaluated at $\phi \left( t+\Delta
t\right) $. (Although the daughter subdomains have slightly varying
$H\left( \phi \right) $ and are not exactly of horizon size, the correction
due to this is negligible.) This gives an integral equation,
\begin{eqnarray}
& \bar{X}\left( \phi _{0}\right)  & =\int d\xi _{1}...d\xi _{N}P\left( \xi
_{1},...,\xi _{N}\right) \nonumber \\ &  & \times \prod
_{i=1}^{N}\bar{X}\left( \phi _{0}+v\left( \phi _{0}\right) \Delta t+\xi
_{i}\sqrt{2D\left( \phi _{0}\right) \Delta t}\right) .\label{eq:xd1}
\end{eqnarray}
Here we use the Ito interpretation of the Langevin equation, where
$v$ and $D$ are evaluated at the initial point of the step,
$\phi =\phi _{0}$.

Since correlations between different daughter subdomains are small
\cite{WV00}, we can approximate the probability density of $\xi _{i}$
by a product of independent identical distributions, $P\left( \xi
_{1},...,\xi _{N}\right) =P\left( \xi _{1}\right) \cdots P\left( \xi
_{N}\right) $. Then Eq.~(\ref{eq:xd1}) becomes
\begin{eqnarray}
&  & \bar{X}\left( \phi _{0}\right) =\nonumber \\
&  & \left[ \int d\xi P\left( \xi \right) \bar{X}\left( \phi _{0}+v\Delta
t+\xi \sqrt{2D\Delta t}\right) \right] ^{\exp \left( 3H\Delta t\right)
}.\label{eq:xd2}
\end{eqnarray}
Rather than trying to solve Eq.~(\ref{eq:xd2}), we consider its
limit for small values of $H\Delta t$. Although the Langevin
description of the random walk given by Eq.~(\ref{eq:lang}) is valid
only for times $H\Delta t\gtrsim 1$, we shall formally consider
it to be valid at all values of $\Delta t$ and take a partial
derivative $\partial /\partial \Delta t$ of Eq.~(\ref{eq:xd2})
at $\Delta t=0$. The same formal limit $\Delta t\rightarrow 0$
is used to derive the FP equations such as Eqs.~(\ref{eq:FP-phys}),
(\ref{eq:FPt}) in the standard stochastic formalism of inflation,
and the same limits of validity apply to the new diffusion equation
{[}Eq.~(\ref{eq:Xequ}){]} that will result from the present argument.

We find
\begin{eqnarray}
&  & \frac{\partial \bar{X}\left( \phi _{0}\right) }{\partial \Delta
t}=0=3H\bar{X}\ln \bar{X}\nonumber \\ &  & +\lim _{\Delta t\to 0}\int d\xi
P\left( \xi \right) \frac{d\bar{X}\left( \phi _{0}+v\Delta t+\xi
\sqrt{2D\Delta t}\right) }{d\phi }\nonumber \\ &  & \times \left(
v+\frac{\xi \sqrt{2D}}{2\sqrt{\Delta t}}\right) .
\end{eqnarray}
We now expand $d\bar{X}/d\phi $ in Taylor series around $\phi =\phi _{0}$
and use the fact that $\xi $ is a normalized random variable,
\begin{equation}
\int d\xi P\left( \xi \right) =1,\quad \left\langle \xi \right\rangle
=0,\quad \left\langle \xi ^{2}\right\rangle =1,
\end{equation}
to obtain an equation for $\bar{X}$,
\begin{equation}
\label{eq:Xequ}
D\frac{d^{2}\bar{X}}{d\phi ^{2}}+v\frac{d\bar{X}}{d\phi }+3H\bar{X}\ln
\bar{X}=0.
\end{equation}

This equation needs to be supplemented with suitable boundary conditions
at end of inflation and/or at Planck boundaries. The probability $\bar{X}$
to have no eternal points for a region which is near the end of inflation
$\phi =\phi _{*}$ should be $1$. Domains reaching Planck
boundaries $\phi _{Pl}$ will never thermalize and we could set
$\bar{X}=0$ at those boundaries; this is consistent with the
viewpoint that super-Planck domains effectively disappear \cite{LLM94}.
Physically meaningful solutions of Eq.~(\ref{eq:Xequ}) should vary
between $0$ and $1$. Therefore, the additional requirements
are
\begin{equation}
\label{eq:XequBC}
0\leq \bar{X}\leq 1;\quad \bar{X}\left( \phi _{*}\right) =1;\quad
\bar{X}\left( \phi _{Pl}\right) =0.
\end{equation}

It is straightforward to verify that Eq.~(\ref{eq:Xequ}), unlike
Eq.~(\ref{eq:FP-phys}), is a gauge-invariant equation. A change
of time variable, $t\to t'$, with $dt'/dt=T\left( t,\phi \right) ,$
will divide the functions $D\left( \phi \right) $ and $v\left( \phi \right)
$, as well as the factor $3H\left( \phi \right) $ in the ``growth''
term, by $T\left( t,\phi \right) $. This extra factor $T\left( t,\phi
\right) $ cancels and Eq.~(\ref{eq:Xequ}) remains unchanged. This is to be
expected since the probability $\bar{X}\left( \phi \right) $
is defined in a gauge-invariant way and should be described by a
gauge-invariant equation.

Equation~(\ref{eq:Xequ}) is a branching diffusion equation similar
to the backward FP equation for the physical volume probability
distribution in the Ito factor ordering, except for the extra factor $\ln
\bar{X}$ that makes it nonlinear. An analogous diffusion equation can be
derived for models of inflation with multiple scalar fields $\phi _{k}$,
\begin{equation}
D\frac{\partial }{\partial \phi ^{k}}\frac{\partial \bar{X}}{\partial \phi
_{k}}+v_{k}\frac{\partial \bar{X}}{\partial \phi _{k}}+3H\bar{X}\ln
\bar{X}=0,
\end{equation}
where
\begin{equation}
v_{k}\left( \phi \right) \equiv -\frac{1}{4\pi }\frac{\partial H}{\partial
\phi ^{k}}
\end{equation}
are the new drift coefficients (in the proper time gauge). Boundary
conditions will be $\bar{X}\left( \phi \right) =1$ along thermalization
boundaries and $\bar{X}\left( \phi \right) =0$ along Planck boundaries
(if any).

\begin{figure}
{\centering \resizebox*{1.5in}{!}{\includegraphics{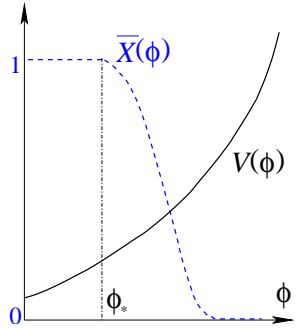}}
\par}

\caption{Probability $ \bar{X}\left( \phi \right) $ to
have no eternal points.}

\label{fig:xplot}
\end{figure}

The qualitative behavior of the probability $\bar{X}\left( \phi \right) $
for a chaotic type inflationary model is illustrated in
Fig.~\ref{fig:xplot}. An approximate WKB solution in the
fluctuation-dominated regime can be obtained if we disregard the drift term
$v\left( \phi \right) \bar{X}'$ and write an ansatz
\begin{equation}
\bar{X}\left( \phi \right) =Ae^{-W\left( \phi \right) },
\end{equation}
where $A$ is approximately constant and $W(\phi )$ is a
slow-changing function, $\left| W''\right| \ll (W')^{2}$ that
satisfies
\begin{equation}
D\left( W'\right) ^{2}-3HW=0.
\end{equation}
The solution is
\begin{equation}
W\left( \phi \right) =6\pi \left[ \int _{\phi _{q}}^{\phi }\frac{d\phi
}{H\left( \phi \right) }\right] ^{2}
\end{equation}
where $\phi _{q}$ is the boundary of fluctuation-dominated range
of $\phi $. It shows that the probability to have eternal points
$1-\bar{X}\left( \phi \right) $ exponentially rapidly approaches
$1$ in the fluctuation-dominated regime.

Since $\bar{X}>0$, there is always a certain (perhaps exceedingly
small) probability for eternal inflation to not be realized within
a given initial horizon-sized region. On the other hand, an inflationary
model could be fine-tuned to entirely disallow eternal inflation,
so that Eqs.~(\ref{eq:Xequ})-(\ref{eq:XequBC}) have no solutions
other than $\bar{X}\equiv 1$. Note that, in the absence of Planck
boundaries, $\bar{X}\equiv 1$ is always a solution of
Eqs.~(\ref{eq:Xequ})-(\ref{eq:XequBC}). This solution would mean that there
are no eternal points anywhere. Equations~(\ref{eq:Xequ})-(\ref{eq:XequBC})
alone do not provide enough information to choose between the constant
solution $\bar{X}\equiv 1$ and a nontrivial solution $\bar{X}\left( \phi
\right) $. However, we know from considerations in Sec.~\ref{sec:ef} that,
if eternal inflation is allowed in a model, then eternal points are
possible, \textit{i.e.}~the probability to have eternal points is nonzero.
Therefore, we should choose the solution $\bar{X}\equiv 1$ only when no
other solution of Eq.~(\ref{eq:Xequ}) that satisfies Eq.~(\ref{eq:XequBC})
can be found.

\subsection{An exact solution}

\begin{figure}
{\centering \resizebox*{3.4in}{!}{\includegraphics{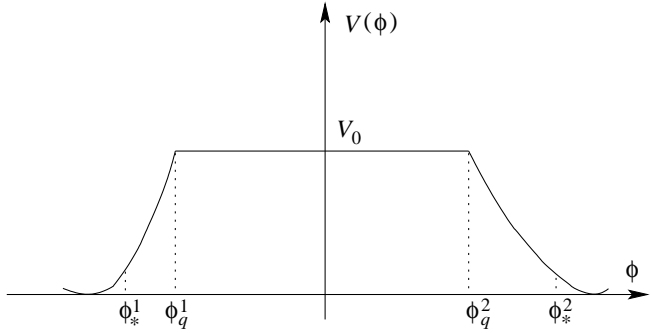}}
\par}
\caption{Illustrative potential for eternal inflation.}

\label{fig:pot2}
\end{figure}

For illustrative purposes, we solve Eq.~(\ref{eq:Xequ}) explicitly
for a flat potential connected to two slow-roll slopes with negligible
diffusion (Fig.~\ref{fig:pot2}). This potential has been considered
in Ref.~\cite{Vil95} as a qualitative illustration of eternal inflation.
The allowed range of $\phi $ is divided into two regimes: in
the fluctuation-dominated regime, $\phi ^{1}_{q}<\phi <\phi ^{2}_{q}$,
the potential is flat and we take $H\left( \phi \right) =H_{0}$,
$D\left( \phi \right) =D_{0}$ and $v\left( \phi \right) =0$.
In the second regime ($\phi _{*}^{1}<\phi <\phi _{q}^{1}$ or
$\phi _{q}^{2}<\phi <\phi _{*}^{2}$), we take $D\left( \phi \right) =0$
and $v\left( \phi \right) \neq 0$, to represent pure deterministic
motion without fluctuations.

We expect that $\bar{X}\equiv 1$ in the second regime, since
all regions thermalize within finite time. In this regime
Eq.~(\ref{eq:Xequ}) becomes
\begin{equation}
v\bar{X}'+3H\bar{X}\ln \bar{X}=0.
\end{equation}
The general solution of this equation is
\begin{equation}
\bar{X}\left( \phi \right) =\exp \left[ -\frac{C}{a\left( \phi \right)
^{3}}\right] ,
\end{equation}
where
\begin{equation}
a\left( \phi \right) \equiv \exp \left[ -4\pi \int _{\phi _{*}}^{\phi
}\frac{Hd\phi }{H'}\right]
\end{equation}
is the scale factor along a slow roll trajectory, and $C$ is
an integration constant. The boundary condition $\bar{X}\left( \phi
_{*}\right) =1$ forces $C=0$ and therefore $\bar{X}\equiv 1$ throughout
the deterministic region, as expected.

\begin{figure}
{\centering \resizebox*{3.4in}{!}{\includegraphics{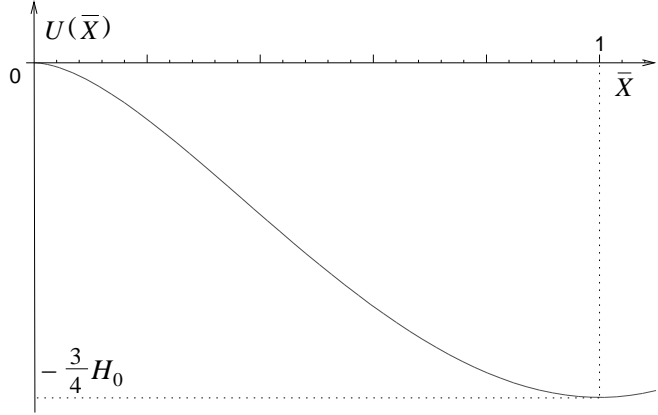}}
\par}

\caption{Effective potential of Eq.~(\ref{eq:U}).}

\label{fig:U}
\end{figure}

Turning now to the fluctuation-dominated regime, we need to solve
\begin{equation}
\label{eq:xd3}
D_{0}\bar{X}''+3H_{0}\bar{X}\ln \bar{X}=0
\end{equation}
with boundary conditions
\begin{equation}
\bar{X}(\phi ^{1,2}_{q})=1.
\end{equation}
For simplicity, we take $\phi _{q}^{2}\equiv \phi _{q}=-\phi _{q}^{1}$.
Equation~(\ref{eq:xd3}) describes a Hamiltonian system of a particle
with mass $D_{0}$ moving in ``time'' $\phi $ in a potential
\begin{equation}
\label{eq:U}
U(\bar{X})=\frac{3H_{0}}{4}\bar{X}^{2}\left( 2\ln \bar{X}-1\right) .
\end{equation}
This potential monotonically decreases from $U=0$ at $\bar{X}=0$
to its minimum $U=-3H_{0}/4$ at $\bar{X}=1$ (see Fig.~\ref{fig:U}).
Boundary conditions correspond to a ``trajectory'' $\bar{X}\left( \phi
\right) $ starting and ending at $\bar{X}=1$ such that the total ``travel
time'' is $2\phi _{q}$. Such a trajectory is unique and is characterized
by the lowest reached value $\bar{X}_{0}$ (the turning point)
such that
\begin{equation}
\label{eq:turn}
\int
_{\bar{X}_{0}}^{1}\frac{dx\sqrt{D_{0}}}{\sqrt{2U(\bar{X}_{0})-2U(\bar{X})}}=\phi _{q}.
\end{equation}
The value $\bar{X}_{0}$ is the lowest probability to have no
eternal points and is naturally achieved at $\phi =0$, in the
middle of the fluctuation-dominated range of $\phi $. In this
model, $\bar{X}_{0}$ is a function of $(\phi _{q}/H_{0})$
which can be obtained numerically. For $\phi _{q}\gtrsim H_{0}$,
one can approximate $U(\bar{X}_{0})\approx 0$ and obtain
\begin{equation}
\bar{X}_{0}\sim 2\exp \left[ -\frac{1}{2}s^{2}-s\right] ,\quad s\equiv 2\pi
\sqrt{3}\frac{\phi _{q}}{H_{0}}\gtrsim 1.
\end{equation}
Numerical evaluation shows that this asymptotic expression holds to
5\% accuracy for $2\pi \phi _{q}/H_{0}>2$ which corresponds to
$\bar{X}_{0}<10^{-4}$. As expected, the probability $X=1-\bar{X}$
to have eternal points is very close to $1$ for regions that
start in the fluctuation-dominated regime.

The opposite case, $\bar{X}_{0}\approx 1$, corresponds to motion
near the minimum of the potential $U(\bar{X})$. Since the period
of such motion is approximately independent of amplitude,
Eq.~(\ref{eq:turn}) cannot be satisfied unless
\begin{equation}
\label{eq:phiqcond}
\phi _{q}>\phi _{c}\equiv \frac{H_{0}}{\sqrt{96}}.
\end{equation}
For $\phi _{q}\leq \phi _{c}$, Eq.~(\ref{eq:xd3}) has no solutions
except $\bar{X}\equiv 1$. This indicates absence of eternal inflation
when the fluctuation-dominated range of $\phi $ is sufficiently
narrow.

It is straightforward to check that for the inflaton potential $V\left(
\phi \right) $ of Fig.~\ref{fig:pot2}, Eq.~(\ref{eq:FP-phys}) with boundary
conditions $P_{V}\left( \phi _{\pm q},t\right) =0$ admits a non-negative
eigenvalue $\gamma _{V}$ if and only if Eq.~(\ref{eq:phiqcond})
holds. In the next subsection we shall see that this result is not
a coincidence.

\subsection{Consistency of conditions for eternal inflation}

\label{sec:cons}We have given two gauge-invariant criteria for presence
of eternal inflation: positivity of the dominant eigenvalue $\gamma _{V}$
of Eq.~(\ref{eq:FP-phys}) and existence of a non-trivial solution
of Eqs.~(\ref{eq:Xequ})-(\ref{eq:XequBC}). Since the relevant diffusion
equations are somewhat dissimilar, a natural question is whether these
two criteria are equivalent. A positive answer is suggested by the
analytic example of the previous section. Here we shall see that
Eqs.~(\ref{eq:Xequ})-(\ref{eq:XequBC}) cannot have a nontrivial solution
$\bar{X}\left( \phi \right) \not \equiv 1$ if the FP equation has a
non-positive dominant eigenvalue.%
\footnote{I do not yet have a proof of existence of a nontrivial solution
$\bar{X}\left( \phi \right) $ when $\gamma _{V}>0$, which is needed to
rigorously demonstrate equivalence of these two criteria.
}

The idea of the proof is to consider a suitable non-negative function
of $\phi $, integrate it over $\phi $ and obtain $\gamma _{V}>0$
as a result.

Let $\mathcal{L}$ be the differential operator of Eq.~(\ref{eq:FP-phys}),
\begin{equation}
\mathcal{L}P_{V}\equiv \frac{\partial ^{2}}{\partial \phi ^{2}}\left(
DP_{V}\right) -\frac{\partial }{\partial \phi }\left( vP_{V}\right)
+3HP_{V}.
\end{equation}
The spectrum of eigenvalues of $\mathcal{L}$ is bounded from
above, and the eigenfunction $\psi _{V}\left( \phi \right) $
of Eq.~(\ref{eq:FPsol}) corresponding to the largest eigenvalue
$\gamma _{V}$ is everywhere positive, as a ground state of a
self-adjoint operator \cite{WV96}. Suppose that a non-trivial solution
$\bar{X}\left( \phi \right) \not \equiv 1$ of
Eqs.~(\ref{eq:Xequ})-(\ref{eq:XequBC}) exists; then the following integral
must be positive,
\begin{equation}
I_{1}\equiv \int _{\phi _{1}}^{\phi _{2}}\left( 1-\bar{X}\left( \phi
\right) \right) \psi _{V}\left( \phi \right) d\phi >0.
\end{equation}
Here $\phi _{1,2}$ are left and right thermalization boundaries,
$\phi _{2}>\phi _{1}$. Since
\begin{equation}
\gamma _{V}=\frac{1}{I_{1}}\int _{\phi _{1}}^{\phi _{2}}\left(
1-\bar{X}\right) \mathcal{L}\psi _{V}d\phi ,
\end{equation}
we would obtain the desired inequality $\gamma _{V}>0$ if we
prove that
\begin{equation}
\label{eq:hitr}
I_{2}\equiv \int _{\phi _{1}}^{\phi _{2}}\left( 1-\bar{X}\right)
\mathcal{L}\psi _{V}d\phi >0.
\end{equation}
Integrating Eq.~(\ref{eq:hitr}) by parts and using boundary conditions
$\bar{X}\left( \phi _{1,2}\right) =1$, we find
\begin{eqnarray}
& I_{2}= & \int _{\phi _{1}}^{\phi _{2}}\psi _{V}\cdot
\mathcal{L}^{*}\left[ 1-\bar{X}\right] d\phi \nonumber \\ &  & +\left.
D\left( \phi \right) \psi _{V}\frac{d\bar{X}}{d\phi }\right| _{\phi
_{1}}^{\phi _{2}}.
\end{eqnarray}
Here $\mathcal{L}^{*}$ is the operator conjugate to $\mathcal{L}$.
Using Eq.~(\ref{eq:Xequ}), we obtain
\begin{eqnarray}
&  & \mathcal{L}^{*}\left[ 1-\bar{X}\right] \nonumber \\
&  & =-D\frac{\partial ^{2}\bar{X}}{\partial \phi ^{2}}-v\frac{\partial
\bar{X}}{\partial \phi }+3H\left( 1-\bar{X}\right) \nonumber \\ &  &
=3H\left( 1-\bar{X}+\bar{X}\ln \bar{X}\right) \geq 0.
\end{eqnarray}
The integral $\int \psi _{V}\mathcal{L}^{*}[1-\bar{X}]d\phi $
is positive because its integrand is nonnegative and not everywhere
zero. The boundary term $D\psi _{V}\bar{X}'$ is non-negative
at $\phi _{2}$ and non-positive at $\phi _{1}$. This proves
Eq.~(\ref{eq:hitr}) and therefore $\gamma _{V}$ must be positive.

It follows that if the dominant eigenvalue $\gamma _{V}$ is zero
or negative, a nontrivial solution $\bar{X}\left( \phi \right) \not \equiv
1$ cannot exist.

\section{Merging of thermalized regions}

\label{sec:global}In the previous section, we found a function $X\left(
\phi \right) $ that describes the probability to realize eternal inflation.
Here we show that the same function gives bounds on probabilities of other
global properties of the set $E$. In particular, we consider
the issue of whether all thermalized domains ultimately merge and
become connected to each other.

\subsection{Global properties of the eternal set}

Observe that the nonlinear diffusion equation for $\bar{X}\left( \phi
\right) $ was derived using the argument that an inflating domain contains
no eternal points if and only if none of its daughter domains, after
one $e$-folding, contain eternal points. It is clear that the
same derivation would apply for the probability $Y\left( \phi \right) $
to realize some other property $Y$ of the set $E$, as long
as that property holds for a given horizon-size inflating domain if
and only if it holds for all daughter domains, and given that the
property $Y$ holds for a thermalized domain ($Y\left( \phi _{*}\right)
=1$). These conditions define a certain class $G$ of ``global''
properties that are binary alternatives, functions of the subset of
$E$ inside a given domain.

We find that the probability to realize any property from this class
$G$ is described by Eqs.~(\ref{eq:Xequ})-(\ref{eq:XequBC}).
The property for the set $E$ to be empty, which occurs with probability
$\bar{X}\left( \phi \right) $, belongs to this class; other examples
would be the property for the set $E$ to not contain any continuous
line segments, or any continuous $2$-dimensional surface segments.
Let us denote by $\bar{S}_{n}$ the property of the eternal set
to contain no fragments of continuous $n$-dimensional submanifolds,
$n=0,1,2,3$.

The universal applicability of Eq.~(\ref{eq:Xequ}) to all global
properties of class $G$ can be interpreted as a statement that
probabilities for global properties of the fractal set $E$ must
be ``fixed points'' under the ``renormalization'' (rescaling)
of the set, and all such fixed points must satisfy Eq.~(\ref{eq:Xequ}).

In presence of eternal inflation, as we have seen,
Eqs.~(\ref{eq:Xequ})-(\ref{eq:XequBC}) admit two solutions, namely
$\bar{X}\equiv 1$ and $\bar{X}=\bar{X}\left( \phi \right) \not \equiv 1$.
Equations~(\ref{eq:Xequ})-(\ref{eq:XequBC}) alone do not give enough
information for us to select one of these two solutions. To find the
probability distribution $Y\left( \phi \right) $ for a particular property
$Y$, other considerations specific to that property and to the given
inflationary model are needed. However, we can conclude \textit{a priori}
that one of these two alternatives must be realized for any given property
$Y$ from the class $G$. This implies that a property from the class $G$
either always holds, $Y\left( \phi \right) \equiv 1$, or holds with
probability $Y\left( \phi \right) =\bar{X}\left( \phi \right) $. (Here
$\bar{X}\left( \phi \right) $ is a fixed function, independent of the
property $Y$.) For instance, it would be enough to show that $Y\left( \phi
\right) >\bar{X}\left( \phi \right) $ for some property $Y$ to conclude
that $Y\left( \phi \right) \equiv 1$.

Since the same function $\bar{X}\left( \phi \right) $ describes
the probability to have no eternal points, a natural question to ask
is whether the occurrence of a given property in a comoving domain
is correlated with presence of eternal points in the same domain.
Consider a property $Y$ of class $G$ and suppose that we
know that $Y$ does not always hold; it follows that the property
$Y$ holds with the probability $Y\left( \phi \right) =\bar{X}\left( \phi
\right) \not \equiv 1$ in a horizon-size domain with value $\phi $ of the
inflaton field. If a horizon-size domain has no eternal points (property
$\bar{X}$), then the property $Y$ must hold in this domain. Therefore the
probability for $Y$ to hold if there are no eternal points is
$Prob(\bar{X}\cap Y)=Prob(\bar{X})=Prob(Y)$, while the probability
for $Y$ to hold in presence of eternal points is $Prob(X\cap
Y)=Prob(Y)-Prob(\bar{X}\cap Y)=0$.

We find that, for any property $Y$ of class $G$, one of
the following two alternatives must hold (in a given inflationary
model): either $Y$ is always true for any horizon-size domain,
or $Y$ is true if and only if the domain contains no eternal
points. In other words, a property $Y$ of class $G$ either
always holds or is perfectly anti-correlated with presence of eternal
points and, by extension, perfectly correlated with other properties
of class $G$. For example, the property $\bar{S}_{0}$ holds
in a domain only when the domain contains no eternal points, while
$\bar{S}_{3}$ always holds. In principle, details of the model
should determine which of these alternatives is realized for a particular
property $Y$.

\subsection{Bounds on merging of thermalized regions}

In this section, we give a bound on merging of thermalized regions
based on the fractal dimension of the eternal set $E$.

The set of all comoving points that will eventually thermalize is
a complement of the set $E$. Thermalized regions will not merge
if the set $E$ encloses pockets of thermalized comoving space.
To enclose a region of a $3$-dimensional space, the set $E$
must contain a continuous boundary of the enclosure which is at least
a $2$-dimensional surface, so the fractal dimension of $E$
must be at least $2$. If the fractal dimension $\textrm{dim}E$
of the set $E$ turns out to be less than $2$, all thermalized
domains must be merged. If, on the other hand, $\textrm{dim}E>2$,
then the question remains unresolved, because there exist sets of
fractal dimension up to $3$ which nevertheless do not enclose
any interior domains.

The condition $\textrm{dim}E<2$ gives a topological bound on
the possibility of merging of thermalized regions. However, this bound
may be rather weak because in typical inflationary scenarios
$\textrm{dim}E=3-\gamma $ with $\gamma \ll 1$. For instance, in a potential
with a flat maximum at $\phi =0$, an estimate of Ref.~\cite{WV96} gives
\begin{equation}
\gamma \approx \frac{1}{16\pi }\frac{\left| V''\left( 0\right) \right|
}{V\left( 0\right) }\ll 1.
\end{equation}
In this case the condition $\gamma >1$ is not satisfied and
the question of merging remains open.

One may distinguish \textit{merging} of all thermalized regions into
one connected domain from \textit{percolation} of thermalized regions,
that is, formation of at least one infinitely large thermalized cluster
(of infinite comoving volume). Percolation is a weaker condition,
since merging entails percolation but not vice versa. Guth and Weinberg
\cite{GW83} have considered the ``old'' scenario of inflation
with bubble nucleation and gave bounds on the nucleation rate $\varepsilon
$ per Hubble $4$-volume for the bubbles (representing thermalized
regions) to percolate. They have found that $\varepsilon \leq 1.1\times
10^{-6}$ is a sufficient condition for non-percolation of bubbles, while
$\varepsilon \geq 0.24$ guarantees percolation. In the model of bubble
nucleation, the fractal dimension of the inflating domain at constant
proper time is \cite{Vil92}
\begin{equation}
\label{eq:dfr-vil}
d_{fr}=3-4\pi \varepsilon /3.
\end{equation}
Since in this model the expansion of inflating domains is homogeneous
with $H=\textrm{const}$, time parametrization by scale factor
is equivalent to proper time parametrization, and the fractal dimension
of the set $E$ has the same value, $\textrm{dim}E=d_{fr}$.
It is interesting to note that the sufficient condition given in
Ref.~\cite{GW83} for percolation of bubbles, $\varepsilon \geq 0.24$,
approximately coincides (to $2$ decimal digits) with the topological bound
$d_{fr}<2$ which guarantees merging.%
\footnote{I am grateful to A. Vilenkin for bringing this point to my
attention. } I conjecture that, in models of scalar field inflation, an
upper bound on the merging probability may be established using the fractal
dimension of the set $E$: namely, merging does not occur if $\textrm{dim}E$
is sufficiently close to $3$.

The property that all regions that eventually thermalized (all points
outside the set $E$) are merged into one connected region is
not a property of class $G$, because merging of thermalized regions
in a domain does not entail merging in all of its subdomains (it is
possible that some subdomains are divided by fragments of eternally
inflating walls which, nevertheless, do not globally enclose an interior
region). We can consider a weaker property that belongs to the class
$G$: the property $\bar{S}_{2}$ for a region to contain
no eternally inflating 2-dimensional surfaces. This property is sufficient
(but not necessary) for merging. On the other hand, if a domain has
a nonzero probability to contain fragments of 2-dimensional eternally
inflating surfaces, it will contain, with the same probability, an
infinite number of arbitrarily small such fragments. If this probability
has been realized, it is likely (but not certain) that merging of
thermalized regions will be impossible in that domain.

From considerations in the previous section we may conclude that,
in a given inflationary model, either the property $\bar{S}_{2}$
always holds (and as a consequence all thermalized domains will always
merge), or the property $\bar{S}_{2}$ never holds in regions
where eternal points are present. This would provide another bound
for the probability of merging if the probability to realize the property
$\bar{S}_{2}$ were determined.

I have been able to show that in any model of scalar field inflation
described by Eq.~(\ref{eq:FP-phys}), the eternal set $E$ cannot
contain any differentiable line segments or fragments of differentiable
surfaces (details of the proof will be given elsewhere). The property
of $E$ to not contain any differentiable line segments also belongs
to the class $G$ and its probability can be shown to exceed $\bar{X}\left(
\phi \right) $; then it follows from considerations of the previous section
that this property must hold for all regions. However, merging of
thermalized domains does not follow from this result, because the fractal
set $E$ may well contain continuous but not differentiable lines
or surfaces.

\section{Summary and discussion}

In this work I have explored the fractal geometry of an eternally
inflating universe at very large scales, in the formalism of stochastic
inflation. I defined a fractal set $E$ consisting of all eternally
inflating points in comoving space. The definition of set $E$
is independent of the choice of spacetime coordinates. Using the
Fokker-Planck equation for comoving volume, I found the fractal dimension
of the set $E$. It coincides with the fractal dimension $d_{fr}$
of the inflating domain as defined in Ref.~\cite{AV87}, if the latter
were computed on surfaces of equal scale factor instead of equal proper
time. In this way, the gauge-invariant construction of the fractal
set $E$ resolves the issue of gauge dependence that plagued previous
calculations of the fractal dimension.

In a recent work, Bousso \cite{Bousso} considered infinite self-reproduction (proliferation) of de Sitter space due to nucleation of black holes. This effect may also create a fractal structure of the spacetime on very large scales. Although the proliferation effect involves topology change which is absent in the usual picture of eternal inflation, the fractal structure of the resulting spacetime can be investigated using the methodology developed above.

I have also examined the conditions for presence of eternal inflation
in a given inflationary model. The original criterion was an unbounded
growth of physical volume of inflating domains, which is equivalent
to the condition $\gamma _{V}>0$ imposed on the largest eigenvalue
$\gamma _{V}$ of the FP equation for the physical volume
{[}Eq.~(\ref{eq:FP-phys}){]}. I have demonstrated that this condition is
gauge-invariant by proving that the sign of the dominant eigenvalue $\gamma
_{V}$ of Eq.~(\ref{eq:FP-phys}) must be the same in all gauges. Then I
found the probability $X\left( \phi \right) $ for an initial horizon-size
region with scalar field value $\phi $ to contain at least one eternal
point. This probability is a solution of a (gauge-invariant) nonlinear
diffusion equation {[}Eq.~(\ref{eq:Xequ}){]} derived in Sec.~\ref{sec:Xequ}
and analyzed in the following sections. To show that the new nonlinear
diffusion equation is consistent with the existing formalism, I checked
that the probability $X\left( \phi \right) $ vanishes when eternal
inflation is not present, in agreement with the criterion $\gamma _{V}>0$.

Finally, I have investigated the issue of merging of thermalized regions
into one connected domain, as observed at arbitrarily late times.
I obtained a topological bound on merging: all thermalized regions
are guaranteed to merge if the fractal dimension of the set $E$
is less than 2. This bound agrees with the previously found sufficient
condition $\varepsilon \geq 0.24$ \cite{GW83} for percolation
in the model of bubble nucleation (in that model, Eq.~(\ref{eq:dfr-vil})
relates the nucleation rate $\varepsilon $ to the fractal dimension
of the set $E$). In Ref.~\cite{GW83}, a sufficient condition
for non-percolation, $\varepsilon \leq 1.1\times 10^{-6}$, was
also found. This suggests a possibility of a stronger bound on merging,
namely that merging does not occur if the fractal dimension is sufficiently
close to $3$.

The stochastic formalism also constrains the probability of merging
in a different way. Since the fractal structure of the eternally inflating
set is preserved on all scales, the probability to realize any global
property of the set must be invariant under rescaling.
Equation~(\ref{eq:Xequ}) expresses the scale invariance of the probability
$\bar{X}\left( \phi \right) $ for a horizon-size region to thermalize, and
the same equation applies to probabilities to realize other global
properties. I have defined a suitable class $G$ of global properties
described by Eq.~(\ref{eq:Xequ}); for example, the property $\bar{S}_{2}$
for a domain to contain no eternally inflating fragments of 2-dimensional
surfaces is a property of class $G$. The property $\bar{S}_{2}$ is a
sufficient condition for merging of thermalized domains (which is itself
not a property of class $G$). I found that the probability to realize
a property of class $G$ is either identically equal to $1$
or is described by the function $\bar{X}\left( \phi \right) $.
This provides a constraint on the probability of merging. However,
Eq.~(\ref{eq:Xequ}) alone does not provide enough information to
resolve this alternative. This is to be expected, because the FP equation
concerns only the total volume of space with given values of $\phi $
and is ignorant of the spatial distribution and shapes of individual
thermalized or inflating domains. One needs to use other considerations,
specific to a particular inflationary model, to select the correct
solution.

Numerical simulations of chaotic inflationary models \cite{LLM94}
suggested that thermalized regions always merge. However, because
of the exponentially growing scales, it is difficult to accurately
simulate the global structure of an eternally inflating spacetime,
especially with realistic model parameters. An unambiguous resolution
of the question of merging in a general scalar-field inflationary
model requires further study.

\begin{acknowledgments}
The author is grateful to Alex Vilenkin for continued encouragement,
fruitful conversations and valuable comments on the manuscript, and
to Xavier Siemens and Vitaly Vanchurin for helpful discussions. This
work was supported by the National Science Foundation.
\end{acknowledgments}

\end{document}